\documentclass[sigplan,twocolumn,nonacm]{acmart}

\usepackage{graphicx}
\usepackage{siunitx}

\usepackage{url}
\usepackage{caption}
\usepackage{subcaption}
\usepackage{enumitem}
\usepackage{tikz}
\usetikzlibrary{arrows.meta, positioning, fit, backgrounds}
\usepackage{listings,listings-rust}

\usepackage[capitalise,nameinlink]{cleveref}
\crefname{algorithm}{Alg.}{Algs.}
\crefname{section}{Sec.}{Secs.}

\RequirePackage{algorithm}
\RequirePackage[noEnd=true]{algpseudocodex}
\algrenewcommand\algorithmicindent{1em}%
\NewCommandCopy{\Oldcomment}{\Comment}
\renewcommand{\Comment}[1]{\Oldcomment{\scriptsize #1}}
\AtBeginEnvironment{algorithmic}{\footnotesize}
\algdef{SL}[REQ]{Req}{0}[1]{requires #1}
\algdef{SL}[ENS]{Ens}{0}[1]{ensures #1}
\algdef{SL}[REPLICASTATE]{ReplicaState}{0}[1]{\textbf{state} #1}
\algdef{SL}[INITIAL]{Initial}{0}[1]{\textbf{initial} #1}
\algdef{SL}[PRE]{Pre}{0}[1]{\textbf{pre} #1}
\algdef{SL}[AWAIT]{Await}{0}[1]{\textbf{await} #1}
\algdef{SL}[ASSERT]{Assert}{0}[1]{\textbf{assert} #1}
\algblockdefx[TYPE]{Type}{EndType}%
[2]{\textbf{type} \emph{#1} #2}{}
\algblockdefx[FN]{Fn}{EndFn}%
[3]{\textbf{fn} \emph{#1} (#2) $\to$ #3}{}
\algblockdefx[PFN]{PubFn}{EndPubFn}%
[3]{\textbf{pub fn} \emph{#1} (#2) $\to$ #3}{}
\algblockdefx[QUERY]{Query}{EndQuery}%
[3]{\textbf{query} \emph{#1} (#2) $\to$ #3}{}
\algblockdefx[UPDATE]{Update}{EndUpdate}%
[2]{\textbf{update} \emph{#1} (#2)}{}
\algblockdefx[MERGE]{Merge}{EndMerge}%
[2]{\hskip\algorithmicindent $\sqcup$ (#1) : #2}{}
\algblockdefx[INVARIANT]{Invariant}{EndInvariant}%
[3]{\textbf{invariant} \emph{#1} (#2) : #3}{}
\algblockdefx[COMPARE]{Compare}{EndCompare}%
[2]{\hskip\algorithmicindent $\sqsubseteq$ (#1) $\to$ #2}{}
\algblockdefx[MUTATE]{Mutate}{EndMutate}%
[3]{\textbf{mutate} \emph{#1} (#2) $\to$ #3}{}
\algblockdefx[INVOKE]{Invoke}{EndInvoke}%
[3]{\textbf{invoke} \emph{#1} (#2) $\to$ #3}{}
\algblockdefx[construct]{Construct}{EndConstruct}%
[3]{\textbf{construct} \emph{#1} (#2) $\to$ #3}{}
\algblockdefx[GENERATE]{Generate}{EndGenerate}%
[1]{\textbf{generate} (#1)}{}
\algblockdefx[EFFECT]{Effect}{EndEffect}%
[1]{\textbf{effect} (#1)}{}
\algblockdefx[PERIODICALLY]{Periodically}{EndPeriodically}%
[0]{\textbf{periodically}}{}
\algblockdefx[ON]{On}{EndOn}%
[1]{\textbf{on} #1}{}
\algtext*{EndUpdate}
\algtext*{EndQuery}
\algtext*{EndGenerate}
\algtext*{EndEffect}
\algtext*{EndFn}
\algtext*{EndPubFn}
\algtext*{EndMutate}
\algtext*{EndInvoke}
\algtext*{EndConstruct}
\algtext*{EndCompare}
\algtext*{EndMerge}
\algtext*{EndInvariant}
\algtext*{EndOn}
\algtext*{EndPeriodically}

\newenvironment{specification}[1][htbp]{%
    \floatname{algorithm}{Spec.}
   \begin{algorithm}[#1]%
  }{\end{algorithm}}

\newcommand{\mvar}[1]{\mathit{#1}}

\newcommand{\mtyp}[1]{\mathrm{#1}}
\newcommand{\mfn}[1]{\mathop{\mathit{#1}}}
\newcommand{\mtd}[1]{\mathit{#1}}

\copyrightyear{2026}
\acmYear{2026}
\setcopyright{cc}
\setcctype{by}
\acmConference[PaPoC '26]{13th Workshop on Principles and Practice of
Consistency for Distributed Data}{April 27, 2026}{Edinburgh,
Scotland}
\acmBooktitle{13th Workshop on Principles and Practice of Consistency for
Distributed Data (PaPoC '26), April 27, 2026, Edinburgh,
Scotland}\acmDOI{00.0000/0000000.0000000}
\acmISBN{000-0-0000-0000-0/0000/00}

\begin{document}

\title{Towards System-Oriented Formal Verification \texorpdfstring{\\}{} of Local-First Access Control}

\author{Florian Jacob}
\email{florian.jacob@kit.edu}
\orcid{0000-0002-5739-8852}
\affiliation{%
  \institution{Karlsruhe Institute of Technology}
  \streetaddress{Kaiserstraße 12}
  \city{Karlsruhe}
  \country{Germany}
  \postcode{76131}
}

\author{Johanna Stuber}
\email{johanna.stuber@student.kit.edu}
\orcid{0000-0000-0000-0000}
\affiliation{%
  \institution{Karlsruhe Institute of Technology}
  \streetaddress{Kaiserstraße 12}
  \city{Karlsruhe}
  \country{Germany}
  \postcode{76131}
}

\author{Hannes Hartenstein}
\email{hannes.hartenstein@kit.edu}
\orcid{0000-0003-3441-3180}
\affiliation{%
  \institution{Karlsruhe Institute of Technology}
  \streetaddress{Kaiserstraße 12}
  \city{Karlsruhe}
  \country{Germany}
  \postcode{76131}
}

\begin{abstract}
Conflict-free replicated data types (CRDTs) and the local-first concept are increasingly employed not only in small-scale collaboration systems among few users who trust each other, but also in large-scale systems, like Matrix for instant messaging and Keyhive for collaborative documents.
Since mutual trust is no longer warranted, these systems require Byzantine fault tolerance and fine-grained access control.
As of today, Matrix and Keyhive pair an informal specification with an unverified reference implementation.
In this work, we follow a bottom-up approach towards constructing formally verified authorization algorithms for Byzantine fault-tolerant local-first systems.
As intermediate target for formal verification, we contribute semantics and invariants of a replicated data type for managing simplified collaboration groups, based on capabilities for access control and hash chronicles for replication.
To enable future integration into local-first systems like Matrix and Keyhive,
we strive for accessibility to system engineers by using
the Rust programming language for formal specification, verification, and implementation, 
enabled by the Verus framework using the Z3 theorem prover at zero runtime cost.
We report on our experience and preliminary results following this approach, and discuss next steps towards scaling up access control expressiveness.
Whether this approach can be scaled up to the complexity of real-world local-first access control systems like Matrix or Keyhive remains future work,
but our findings demonstrate the potential of system-oriented formal verification with Verus.
\end{abstract}



\maketitle

\newpage
\section{Introduction} 
\label{sec:intro}

Conflict-Free Replicated Data Types (CRDTs,~\cite{shapiroConflictFreeReplicatedData2011,almeida_approaches_2024}) are a common solution to concurrency conflicts in local-first systems, i.e., decentralized systems designed for low latency and availability under partition~\cite{kleppmann_local-first_2019,kleppmannCritiqueCAPTheorem2015}.
CRDT-based systems do not attempt to conceal their inherent concurrency,
but instead demand concurrent thinking from their designers.
In the CRDT research community, the benefits of formal methods for CRDT design to ensure consistency and application invariants are well-established, especially in Byzantine contexts~\cite{gomes_verifying_2017,da_design_2024}, to counter the highly-concurrent and nondeterministic nature of distributed systems.
However, Basin et al. recently called attention to the gap between formal methods theory and distributed systems practice relying on informal specifications and unverified reference implementations~\cite{basin_it_2025}.
An example for this gap is the Matrix decentralized system for instant messaging and group chats~\cite{matrix_spec_v1.16},
widely used in security-critical settings ranging from the French public sector and the German healthcare system to the United Nations and NATO~\cite{element_software_gmbh_matrix_2025,elementsoftwaregmbhNATONI2CEMessenger2024}.
A recent update of the Matrix specification dealt with an algorithm violating a security invariant~\cite{dougal_hydra_2025,matrix_rooms_v12},
which could have been caught early by integrating formal methods into the standardization process.

Large-scale local-first systems like Matrix cannot rely on mutual trust but face arbitrary malevolence of participants, necessitating Byzantine fault tolerance and fine-grained access control.
We therefore study formal verification of local-first access control in an asynchronous system with Byzantine faults.
We strive for accessibility to system engineers by using Verus~\cite{lattuada_verus_2023,lattuada_verus_2024}, a recent verification framework that
pairs the popular high-performance programming language Rust with the Z3 theorem prover at zero runtime cost. 

Instead of attempting to verify the authorization algorithms of Matrix or Keyhive directly, we follow a bottom-up approach to construct simple but formally verified authorization algorithms for local-first systems, in order to gradually extend complexity and expressiveness in the future.
We contribute a semantics specification of a replicated data type for capability-based access control in collaboration groups, along with a simplified but verified algorithm for local-first authorization.
We report on our experience with Verus to unify specification, implementation, and verification in a single Rust source file, and publish all source code under a free license~\cite{mftf-source}.

\section{Local-First Systems and Related Work}

Inspired by local-first software~\cite{kleppmann_local-first_2019,kleppmannPresentFutureLocalfirst2024}, we use the term \emph{local-first} to describe decentralized systems where every entity independently maintains its own local replica of system state and acts autonomously, i.e., autonomously executes query and mutate operations. 
Instead of necessitating up-front coordination among entities,
information on state changes is exchanged asynchronously.
Thus, the operations offered by an entity are available with local latency regardless of network partitions and faults in other entities.
In addition, large-scale local-first systems demand algorithms that tolerate Byzantine faults, since trust among entities is no longer warranted at scale.
For fault tolerance, low latency, and eventual consistency, local-first systems rely on previous work on wait-freedom~\cite{herlihyWaitfreeSynchronization1991}, coordination avoidance~\cite{bailisCoordinationAvoidanceDatabase2014}, and conflict-free replicated data types~\cite{shapiroConflictFreeReplicatedData2011,almeida_approaches_2024}.

A \emph{group} is what we call a replicated object together with the set of entities that replicate and collaborate on that object,
i.e., a chat group or collaborative document.
The group's state encompasses both data, like the document state, and metadata, like the group name or permission assignments.
Group state is replicated via a partially ordered set of state change \emph{events} using the \emph{hash chronicle} data type~\cite{jacob_best_2025}.
Hash chronicles are Byzantine fault-tolerant CRDTs that replicate a partially ordered event set via recursive hash linking:
Any event is bound to its \emph{causal history}, i.e., the chronicle subset that contains all causal precursor events back to group creation, by recursively linking its direct causal precursors.
The concept behind hash chronicles represents the current community consensus for replication in large-scale local-first systems; the concept is also known as hash-linked DAGs (directed acyclic graphs, ~\cite{kleppmann_making_2022}), blocklace~\cite{almeida_blocklace_2025}, Matrix Event Graph~\cite{jacobAnalysisMatrixEvent2021}, or MerkleDAG~\cite{sanjuanMerkleCRDTsMerkleDAGsMeet2020}.

Among the most sophisticated examples of large-scale local-first systems and frameworks is the Matrix~\cite{matrix_spec_v1.16} group communication system, as well as the Automerge/Keyhive ~\cite{automergecontributorsAutomergeVersionControl2026,zelenkaKeyhiveLocalfirstAccess} and p2panda~\cite{p2pandacontributorsP2panda2026} CRDT frameworks for local-first collaborative applications.
While Matrix 
is widely deployed already,
Keyhive and p2panda are 
still in the research project phase, but also aim to scale collaborative local-first systems up to, e.g., a decentralized Wikipedia equivalent~\cite{zelenkaKeyhiveLocalfirstAccess}.


We recently proposed a conceptual model and security properties for a Matrix-like local-first authorization algorithm in an informal top-down approach~\cite{jacob_best_2025}.
In this paper, we follow a bottom-up approach and focus on constructing local-first access control semantics and simplified but formally verified algorithms,
instead of characterizing what is currently present in Matrix. 
In his work on local-first authorization, Kleppmann focuses on resolving mutual authorization revocations among equally-privileged entities~\cite{kleppmann_papoc_keynote_2025}, which are out of scope for this work.
The work of Rault et al. explores CRDT-based access control under the honest-but-curious assumption instead of Byzantine faults, and focuses on delegation, revocation, and compensating actions consecutive to access control concurrency conflicts ~\cite{raultAccessControlBased2023,rault_access_2024}.
Other work on local-first authorization usually has different assumptions on Byzantine faults~\cite{rault_access_2024,weberAccessControlWeakly2016},
while formal work on CRDTs~\cite{almeida_approaches_2024} usually does not integrate authorization~\cite{da_design_2024,gomes_verifying_2017}.



\begin{figure}[t]
\centering
    \begin{subfigure}[t]{0.4\linewidth}
        \vspace{0em}
        (i)
        \vspace{-1em}
        
        \includegraphics[width=\linewidth]{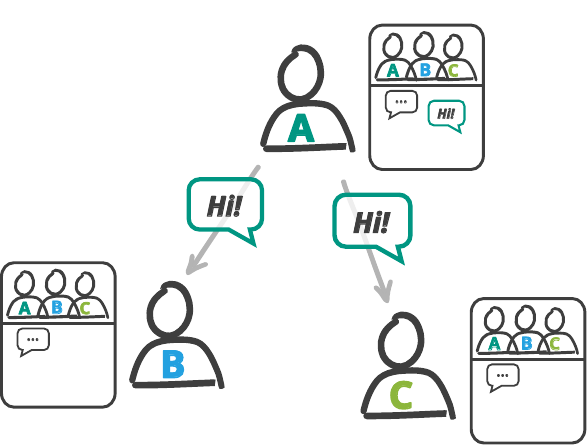}
    \end{subfigure}
    \hspace{2em}
    \begin{subfigure}[t]{0.4\linewidth}
        \vspace{0em}
        (ii)
        
        \vspace{-1em}
        \includegraphics[width=\linewidth]{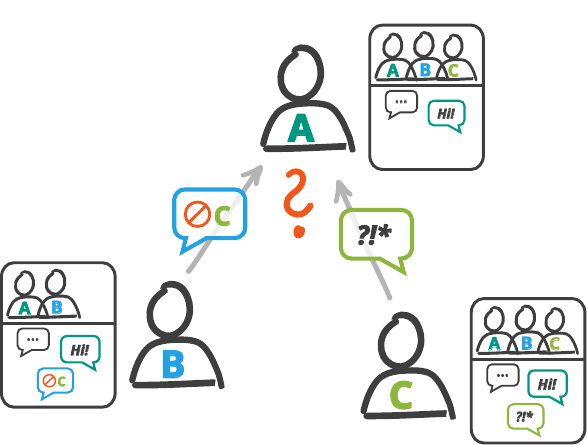}
    \end{subfigure}
    \caption{
        A chat group consisting of entities $A, B, C$.
        (i) Entity $A$ adds a new message to its local chat history first, then broadcasts to $B$ and $C$.
        (ii) Entity $C$ adds a message while concurrently, $B$ revokes the authorization that $C$ utilized.
    }
    \label{fig:conflicts}
\end{figure}

\begin{figure}[t]
\centering
    \includegraphics[width=.8\linewidth,trim={5cm 3cm 4.7cm 4.2cm},clip]{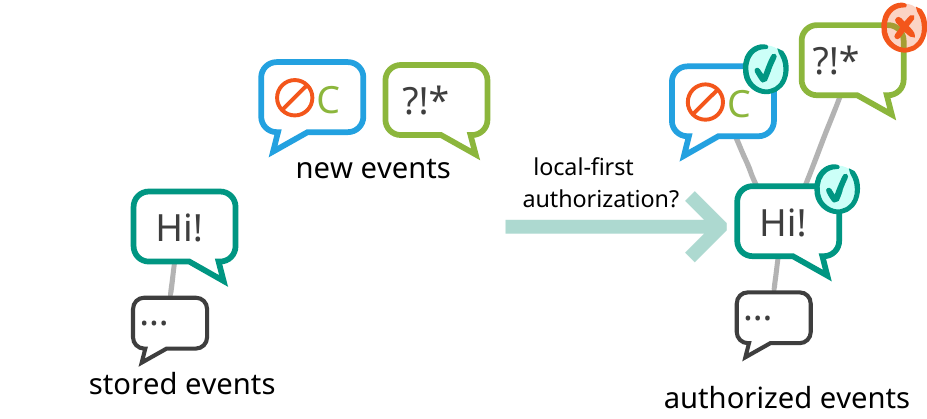}
    \caption{
        Given new events and stored events only,
        specifications for local-first authorization define how entities independently decide authorization of each event.
    }
    \label{fig:lf-auth}
\end{figure}

%
%

\section{Problem Specification}
As target for formal verification, pursued but not yet met in~\cref{sec:formal-verification}, we now specify the semantics and invariants of a Byzantine fault-tolerant, replicated data type for managing collaboration groups (cf. \cref{fig:conflicts}) via access control capabilities.
While we focus on administrative access rights, i.e., rights to grant or revoke capabilities, our type also exemplarily supports querying and mutating a group name.

At the example of assigning a new group name,
invocations, mutate, and query operations act together as follows:
First, an invocation $v \in \mathbb{V}_\mtyp{assign}$ is constructed, supplying the new group name.
Second, the invocation is logged in the group chronicle via its $\mfn{log}(v)$ mutate operation, which combines the invocation with a timestamp to form an event.
The group chronicle represents a replicated log of events, partially ordered by timestamp.
For Byzantine fault tolerance, an event's timestamp is the set of identifier hashes of all its precursor events, instead of, e.g., the usual version vector~\cite{almeida_approaches_2024}.
Chronicle reconciliation, as described e.g. in~\cite{jacob_logical_2024}, provides eventual consistency based on the $\mfn{join}()$ mutate operation that merges two potentially diverging group chronicles to their least upper bound.
Executing the $\mfn{values}()$ query operation results in the set of latest, authorized $\mtyp{assign}$ events.

In access control terminology,
a capability is one way to encode the access right (authorization) of a subject entity to invoke an action (operation) on a replicated object, potentially affecting a specific object entity or event.
As shown in \cref{fig:lf-auth}, the authorization algorithm gets the entity's current events and new events as input, and decides on the authorization of each event as output.
The challenge of local-first authorization is to specify authorization semantics and security invariants that do not require coordination or trust among entities, but yet are as expressive as possible.

\paragraph{System Assumptions and Threat Model}
We assume that an arbitrary strict subset $f$ of the $n$ overall system entities is Byzantine, i.e., $f/n < 1$.
Byzantine entities may deviate arbitrarily from system algorithms, which includes malicious intent.
The following assumptions restrict the abilities of Byzantine entities:
Events are authenticated and integrity-protected, e.g. by digital signatures.
Further, the access control system assumes a Byzantine fault-tolerant chronicle implementation that provides eventual consistency~\cite{jacob_logical_2024}.
The replicated chronicle is assumed to be \emph{append-only} for correct and Byzantine entities alike, achievable in practice through hash linking:
any event includes the hashes of its precursor events (cf.~\cite{jacob_logical_2024}).
Thereby, with a collision-resistant hash function, the causal history of any event is immutable: an attacker cannot forge new precursors of existing events.
In practice, \emph{recursive} hash linking allows to reduce timestamps to the hashes of the direct precursor events.

Byzantine entities have three main attack vectors: \emph{omission}, \emph{equivocation}, and \emph{backdating}~\cite{jacob_logical_2024}.
\emph{Omission} allows Byzantine entities to selectively decide with which entities to share events with.
Byzantine omission can be mitigated by simple anti-entropy mechanisms, as long as vector clocks and similar mechanisms relying on honest entities are avoided.
\emph{Equivocation} allows Byzantine entities to create two successor events concurrent to each other, while correct entities would usually assign an ordering between successor events.
Equivocation aims to make correct entities receive events in a different ordering to exploit ordering dependencies in event processing.
\emph{Backdating} allows Byzantine entities to claim subsets of the chronicle as causal history, while correct entities would usually append events following the most recent events in their local state.
Backdating can be seen as a generalization of equivocation.
However, backdating aims to make new events look like events that were not received for a long time, to exploit the mechanisms that ensure order-independent event processing.
While eventual consistency ensures that all correct entities eventually know of all events from equivocation and backdating, the access control system still has to decide how to deal with these events in an eventually consistent way.
A secure local-first access control system must prevent Byzantine influence on the safety of event authorization, capability revocation, and query results through backdating and equivocation.

\paragraph{Data Type Semantics}
\label{sec:semantics}
We discuss the semantics and invariants in mathematical form to define the intermediate goal pursued in formal verification.
The Rust formalization discussed in~\cref{sec:formal-verification} consists of different simplifications of this data type.
We define the group type as a composition of the underlying group chronicle and types for the group's capabilities and name.
We define query operations for capabilities, authorization, and group name based on local group state,
as well as
operation invocations for granting and revoking capabilities, and assigning a group name.
Invocations mutate group state by being logged to the group chronicle.
We assume that the group chronicles is replicated as Byzantine fault-tolerant hash chronicle.
Inspired by Keyhive~\cite{zelenkaKeyhiveLocalfirstAccess}, we define authorization semantics based on capabilities~\cite{sandhuAccessControlPrinciple1994,gusmeroliCapabilitybasedSecurityApproach2013}:
Authorized subject entities can grant an object entity the capability to invoke a specific operation, encoded as a capability grant event in the chronicle.
As claim of authorization, invocations of authorized subject entities present the identifier of a corresponding capability grant.
Authorized subject entities can revoke a capability grant again, also referencing the grant event identifier.
We believe that capabilities in this “granted-present-revoke” style, analogous to “observed-remove” set CRDTs~\cite{almeida_approaches_2024}, are well-suited to capture user intent in a way supported by local-first authorization algorithms, while resulting in desirable conflict resolution semantics~\cite{weidnerDesigningDataStructures2022}.

\begin{specification}[t]
\caption{A group chronicle provides query and mutate operations on its partially ordered log of events.
Events combine an invocation with a timestamp. A timestamp is a set of precursor event identifiers. Identifiers are collision-resistant.}
\label{alg:chronicle-model}
\begin{algorithmic}

\Type{Event}{$\mathbb{E} \subseteq \mathbb{T} \times \mathbb{V}$} \Comment{an event combines a timestamp with an invocation}
\Construct{event}{$T \in \mathbb{T}, v \in \mathbb{V}$}{$e \in \mathbb{E}$}
\EndConstruct
\Query{tme}{$e \in \mathbb{E}$}{$T \in \mathbb{T}$}
\EndQuery
\Query{voc}{$e \in \mathbb{E}$}{$v \in \mathbb{V}$}
\EndQuery
\Query{$<$}{$e_1, e_2 \in \mathbb{E}$}{$\mfn{id}(e_1) \in e_2.\mtd{tme}$}
\EndQuery
\Query{$\parallel$}{$e_1, e_2 \in \mathbb{E}$}{$e_1 \neq e_2 \land e_1 \nless e_2 \land e_2 \nless e_1$}
\EndQuery
\EndType
\Type{EventIdentifier}{$\mathbb{I}$} \Comment{identifiers provide unique event representations.}
\Construct{id}{$e \in \mathbb{E}$}{$i \in \mathbb{I}$}
\Ens $\forall e_1, e_2 \in \mathbb{E}\colon \mfn{id}(e_1) = \mfn{id}(e_2) \Leftrightarrow e_1 = e_2$ \Comment{collision resistance}
\EndConstruct
\EndType
\Type{Timestamp}{$\mathbb{T} \subseteq \mathcal{P}(\mathbb{I})$} \Comment{downward-closed sets of event identifiers}
\Construct{from}{$E \subseteq \mathbb{E}$}{$\{ \mfn{id}(e) \mid e \in E \}$}
\EndConstruct
\Query{valid}{$T \in \mathbb{T}$}{$a \in \{\bot, \top\}$} \Comment{$T$ downward-closed?}
\Statex \hspace{2em} $a \gets \forall i \in T, e, \check{e} \in \mathbb{E}\colon i = \mfn{id}(e) \land \check{e} < e \Rightarrow \mfn{id}(\check{e}) \in T$
\EndQuery
\EndType

\Type{GroupChronicle}{$\mathbb{G} \subseteq \mathcal{P}(\mathbb{E})$}
\Comment{group chronicles are sets of events}
\Construct{$\mvar{create}$}{$ \mvar{sbj} \in \mathbb{S} $}{$v \in \mathbb{V}_\mtyp{create} $} \Comment{declares initial group setup}
\EndConstruct
\Mutate{log}{$ G \in \mathbb{G}, v \in \mathbb{V} $}{$g' \in \mathbb{G}$}
\Statex \hspace{2em} $G' \gets G \cup \{ \mfn{event}(G.\mtd{now}(), v) \}$ \Comment{logs invocation $v$ as event at $G.\mtd{now}()$}
\EndMutate
\Mutate{join}{$ G_1 \in \mathbb{G}, G_2 \in \mathbb{G} $}{$G' \in \mathbb{G}$}
\Statex \hspace{2em} $G' \gets G_1 \cup G_2 $ \Comment{merges $G_1$ and $G_2$ to their least upper bound}
\EndMutate

\Query{events}{$ G \in \mathbb{G} $}{$ E \subsetneq \mathbb{E}$}
\Statex \hspace{2em} $E \gets \{e \in G \}$
\EndQuery
\Query{now}{$ G \in \mathbb{G} $}{$ T \in \mathbb{T}$} \Comment{current timestamp of $G$}
    \Statex \hspace{2em} $T \gets \mathbb{T}.\mtd{from}(G.\mtd{events}())$
\EndQuery
\Query{$\mvar{creation}$}{$G \in \mathbb{G}$}{$e_c \in G \cap \mathbb{E}_\mtyp{create}$} \Comment{create event of $G$}
  \Req $ G.\mtd{valid}() $ 
\EndQuery
\Query{pre}{$G \in \mathbb{G}, e \in \mathbb{E}$}{$\check{G} \subseteq G$} \Comment{wind $G$ back precursors of $e$}
  \Req $ G.\mtd{valid}() \land e \in G $
  \Statex \hspace{2em} $ \check{G} \gets \{ \check{e} \in G \mid \check{e} < e \} $
\EndQuery
\Query{conc}{$G \in \mathbb{G}, e \in \mathbb{E}$}{$\check{G} \subseteq G$} \Comment{precursor and events concurrent to $e$}
  \Req $ G.\mtd{valid}() \land e \in G $
  \Statex \hspace{2em} $ \check{G} \gets \{ \check{e} \in G \mid \check{e} < e \lor \check{e} \parallel e \} $
\EndQuery
\Query{valid}{$ G \in \mathbb{G} $}{$ a \in \{ \bot, \top \} $}
    \Statex \hspace{2em} $ a \gets \exists_{=1} e_c \in G \cap \mathbb{E}_\mtyp{create} $ \Comment{create event is unique}
    \Statex \hspace{3em} $ \land \forall e \in G\colon e < e_c \lor e = e_c \lor e > e_c $
    \Comment{no events concurrent to $e_c$}
    \Statex \hspace{3em} $ \land \mvar{e.tme}.\mtd{valid}() $ \Comment{timestamp is downward-closed}
\EndQuery


\EndType
\vspace{-1.4em}
\end{algorithmic}
\end{specification}

\begin{specification}[t]

\caption{The group capabilities type extends the chronicle with capability management and queries for authorization. 
Assigning the group name requires the corresponding capability.
Boxes show authorization algorithm calls.}
\label{alg:acs-model}
\begin{algorithmic}
\Type{GroupCapabilities}{$C(\mathbb{G})$} \Comment{group capabilities are sets of $\mtyp{grant}$ events}

\Construct{grant}{$\mvar{sbj} \in \mathbb{S}, \mvar{grnt} \in \mathbb{I}, \mvar{cap} \in \mathbb{M}, \mvar{obj} \in \mathbb{S}$}{$v \in \mathbb{V}_\mtyp{grant}$} 
\EndConstruct
\Construct{revoke}{$\mvar{sbj} \in \mathbb{S}, \mvar{grnt} \in \mathbb{I}, \mvar{obj} \in \mathbb{I}$}{$v \in \mathbb{V}_\mtyp{revoke}$}
\EndConstruct

\Statex

\Query{$\mvar{caps}$}{$G \in \mathbb{G}$}{$\mvar{cs} \subseteq G \cap \mathbb{E}_\mtyp{grant}$} \Comment{granted, unrevoked capabilities}
  \Req $ G.\mtd{valid}() $
  \Statex \hspace{2em} $ \mvar{cs} \gets \{ \mvar{gr} \in G \cap \mathbb{E}_\mtyp{grant} \mid \:\BoxedString{G.\mtd{authorizes}(\mvar{gr})} \; $
  \Statex \hspace{3.15em} $ \land (\nexists \mvar{rv} \in G \cap \mathbb{E}_\mtyp{revoke}\colon \BoxedString { G.\mtd{authorizes}(\mvar{rv}) } \land \mvar{rv.voc.obj} = \mfn{id}(\mvar{gr}) ) \} $
\EndQuery

\Query{\BoxedString{$\mvar{authorizes}$}}{$G \in \mathbb{G}, e \in \mathbb{E}$}{$a \in \{ \bot, \top \}$} \Comment{$e$ authorized by $G$?}
  \Req $ G.\mtd{valid}() $
  \Req $ \mvar{e.tme} \subseteq G.\mtd{now}() \land \mvar{e.tme}.\mtd{valid}()$ 
  \Statex \hspace{2em} $ a \gets e = G.\mtd{creation}() $ \Comment{creation is authorized by definition}
  \Statex \hspace{3em} $ \lor e < G.\mtd{creation}() $ \Comment{creation authorizes its precursors}
  \Statex \hspace{3em} $ \lor ($ \Comment{other events need a matching, authorized grant precursor}
  \Statex \hspace{4.6em} $ \exists \mvar{gr} \in G.\mtd{pre}(e) \cap \mathbb{E}_\mtyp{grant}\colon G.\mtd{authorizes}(\mvar{gr}) $
  \Statex \hspace{11em} $ \land \mfn{id}(gr) = \mvar{e.voc.grnt} $
  \Statex \hspace{11em} $ \land \mvar{e.voc.sbj} = \mvar{gr.voc.obj} \land \mvar{e.voc} \in \mathbb{V}_\mvar{gr.voc.cap} $
  \Statex \hspace{4em} $ \land \nexists \mvar{rv} \in G.\mtd{conc}(e) \cap \mathbb{E}_\mtyp{revoke}\colon G.\mtd{authorizes}(\mvar{rv})$
  \Statex \hspace{11em} $ \land \mvar{rv.voc.obj} = \mvar{e.voc.grnt} $
  \Statex \Comment{…but no matching, authorized revocation, before or concurrent}
  \Statex \hspace{3em} $ ) \land e \in \mathbb{E}_\mtyp{revoke} \Rightarrow \exists gr \in G.\mtd{pre}(e) \cap \mathbb{E}_\mtyp{grant}\colon \mfn{id}(gr) = \mvar{e.voc.obj} $
  \Statex \Comment{revocations must match a grant precursor}
\EndQuery



\EndType


\Type{GroupName}{$N(\mathbb{G})$} \Comment{group name is the latest set of $\mtyp{assign}$ events}

\Construct{$\mvar{assign}$}{$\mvar{sbj} \in \mathbb{S}, \mvar{grnt} \in \mathbb{I}, n \in N $}{$v \in \mathbb{V}_\mtyp{assign}$}
\EndConstruct

\Query{$\mvar{values}$}{$g \in \mathbb{G}$}{$\mvar{vs} \in \mathcal{P}(\mathbb{E}_\mtyp{assign})$} \Comment{latest group names}
  \Req $G.\mtd{valid}()$
  \Statex \hspace{2em} $\mvar{vs} \gets \{ \hat{e} \in g \cap \mathbb{E}_\mtyp{assign} \mid \, \BoxedString{ g\!\mfn{.authorizes}(\hat{e}) } \, $ \Comment{latest, authorized assigns}
  \Statex \hspace{3.6em} $ \land (\nexists e \in g \cap \mathbb{E}_\mtyp{assign}\colon \BoxedString{ g.\mtd{authorizes}(e) } \; \land e > \hat{e} ) \}$
\EndQuery
\EndType
\vspace{-1.4em}

\end{algorithmic}
\end{specification}

Spec.~\ref{alg:chronicle-model} contains the chronicle type as eventually consistent log of events, partially ordered by timestamps, interlinked by event identifiers.
Spec.~\ref{alg:acs-model} extends the group chronicle with capabilities and authorizations, as well as a group name.

In Spec.~\ref{alg:chronicle-model},
an event $e \in \mathbb{E}$ encodes the invocation $\mvar{e.voc} \in \mathbb{V_\mtyp{m}}$ of an operation, like a group name assignment, at time $\mvar{e.tme} \in \mathbb{T}$.
An event $e$ has a unique identifier $\mfn{id}(e) \in \mathbb{I}$ used to reference the event, i.e., a unique digest from a collision-resistant hash function.
To ease notation, a timestamp $T \in \mathbb{T} \subsetneq \mathcal{P}(\mathbb{I})$ represents the causal history of an event as the downward-closed set of identifiers of all precursor events.
In practice, timestamps would be compressed to the identifier set of the direct precursors.
Invocations have a subtype in $\mathbb{M} = \{ \mtyp{create}, \mtyp{grant}, \mtyp{revoke}, \mtyp{assign} \}$, we write $\mathbb{V}_{m \in \mathbb{M}}$ for all invocations of subtype $m$, and $\mathbb{E}_{m \in \mathbb{M}}$ for all events with invocations of subtype $m$.
The group create event and its precursors, used for initial group state setup, are authorized by definition, other events $e$ present a reference to a capability grant event, $\mvar{e.voc.grnt}$, as claim of authorization.
A group state $G$ is a set of events $G \in \mathbb{G} \subseteq \mathcal{P}(\mathbb{E})$, partially ordered by their timestamps.

The chronicle type provides the following operations:
\begin{description}
    \item[$\mfn{create}(\mvar{sbj} \in \mathbb{S}) \to v \in \mathbb{V}_\mtyp{create}$] constructs a group $\mtyp{create}$ invocation, to be logged after a set of invocations that define initial group state.
    Logging a $\mtyp{create}$ event marks the end of group setup by making the chronicle valid.
    \item[$\mfn{log}(G \in \mathbb{G}, v \in \mathbb{V}) \to G' \in \mathbb{G}$] logs invocation $v$ as successor event of all events in $G$.
    \item[$\mfn{join}(G_1 \in \mathbb{G}, G_2 \in \mathbb{G}) \to G' \in \mathbb{G}$] merges two diverging chronicles to their least upper bound via set union, providing the foundation for eventual consistency.
    \item[$\mfn{events}(G \in \mathbb{G}) \to E \in \mathbb{E}$] returns all events in $G$.
    \item[$\mfn{now}(G \in \mathbb{G}) \to T \in \mathbb{T}$] derives the current timestamp considering all events in $G$.
    \item[$\mfn{creation}(g \in \mathbb{G}) \to e_c \in \mathbb{E}_\mtyp{create}$] returns the $\mtyp{create}$ event.
    \item[$\mfn{pre}(G \in \mathbb{G}, e \in \mathbb{E}) \to \check{G} \in \mathbb{G}$] winds chronicle $G$ back to the state $\check{G}$ right before $e$ was logged.
    \item[$\mfn{conc}(G \in \mathbb{G}, e \in \mathbb{E}) \to \check{G} \in \mathbb{G}$] winds $G$ back to $\check{G}$, including only precursors to and events concurrent to $e$.
    \item[$\mfn{valid}(G \in \mathbb{G}) \to a \in \{\bot, \top\}$] verifies that $G$ has a unique create event, that there is no event concurrent to the create event, and that all events have a valid, i.e., downward-closed, timestamp.
\end{description}

Spec.~\ref{alg:acs-model} extends the group chronicle with operations for managing group capabilities and group name.
Capability-based access control is provided by the following operations:

\begin{description}
    \item[$\mfn{grant}(\mvar{sbj} \in \mathbb{S}, \mvar{grnt} \in \mathbb{I}, \mvar{cap} \in \mathbb{M}, \mvar{obj} \in \mathbb{S}) \to v \in \mathbb{V}_\mtyp{grant} \hspace{2em}$]
    constructs an invocation in which entity $\mvar{sbj}$ grants an entity $\mvar{obj}$ the capability to invoke operations corresponding to $\mvar{cap}$. The invocation presents $\mvar{grnt}$ as claim of authorization.
    \item[$\mfn{revoke}(\mvar{sbj} \in \mathbb{S}, \mvar{grnt} \in \mathbb{I}, \mvar{obj} \in \mathbb{I}) \to v \in \mathbb{V}_\mtyp{revoke}$] constructs an invocation in which entity $\mvar{sbj}$, revokes the previous capability grant event referenced in $\mvar{obj}$, presenting $\mvar{grnt}$ for authorization.
    \item[$\mfn{caps}(G \in \mathbb{G}) \to \mvar{cs} \in \mathcal{P}(\mathbb{E}_\mtyp{grant})$] returns the set of authorized $\mtyp{grant}$ events in $G$ that have not been revoked yet.
    \item[$\mfn{authorizes}(G \in \mathbb{G}, e \in \mathbb{E}) \to \mvar{a} \in \{\bot, \top\}$] whether group state $G$ authorizes event $e$, mainly verifying that the precursors of $e$ must contain a matching, authorized grant; while precursors and concurrent events must not contain a matching, authorized revocation.
\end{description}

The group name is managed by the following operations:

\begin{description}
    \item[$\mfn{assign}(\mvar{sbj} \in \mathbb{S}, \mvar{grnt} \in \mathbb{I}, n \in \mathbb{N}) \to v \in \mathbb{V}_\mtyp{assign}$] constructs an invocation in which entity $\mvar{sbj}$ assigns name $n$, claiming authorization by $\mvar{grnt}$.
    \item[$\mfn{values}(G \in \mathbb{G}) \to \mvar{vs} \in \mathcal{P}(\mathbb{E}_\mtyp{assign})$] returns the set of the most recent, authorized $\mtyp{assign}$ events in $G$, i.e., using multi-value resolution of concurrency conflicts.
\end{description}

In the remainder of this section, we discuss the security invariants underlying our design.
The semantics of a capability grant event is to authorize successor events presenting the granted capability, while a revocation event matching the grant deauthorizes successor and concurrent events of the revocation that present the granted capability.
It seems consequential that the authorization of any event must thereby depend on precursor and concurrent events only, and that successor events must not affect authorization of precursors.
However, note that this is a fallacy induced by concurrency seeming like a transitive relation especially in graphical representations, while, in fact, it is non-transitive:
Assume a revocation $a_i$ de-authorizes a concurrent usage $b_i$.
Then, a revocation $c_{i+1}$ concurrent to $a_i$ but successor to $b_i$ de-authorizes $a_i$ and thereby re-authorizes $b_i$, despite being a precursor of $c_{i+1}$.
This anomaly follows from the expansion of the scope of effect of revocations to concurrent events.
However, the expansion is necessary to ensure safety of revocations against backdating:
Byzantine entities can backdate to exclude a revocation from the precursors of new events, in an effort to circumvent the revocation.
Therefore, to make backdating ineffective, authorized revoke events must affect authorization of successors as well as concurrent events~\cite{kleppmann_papoc_keynote_2025}.

Related work differentiates between two notions of event authorization, which we call \emph{precursive} authorization and \emph{discursive} authorization (cf. storage and execution authorization in~\cite{jacob_best_2025}).
These notions do not define \emph{how} authorization is performed, but differentiate subsets of group state on which authorization is based on.
%
%
%
\emph{Precursive authorization} is based on the precursors of event $e$ in group $G$, i.e., given by $G.\mtd{pre}(e).\mtd{authorizes}(e)$.
As the precursor set is immutable, precursive authorization is immutable as well.
Correct entities broadcast and merge precursively authorized events only.
\emph{Discursive authorization}
is based on precursors and events concurrent to $e$, given by $G.\mtd{authorizes}(e)$.
However, as discursive authorization of concurrent events depends on \emph{their} concurrent events, any event in $G$ may affect discursive authorization of $e$, regardless of the relation to $e$.
Thereby, discursive authorization is mutable: events may lose or regain discursive authorization due to new revocation events.

We characterize invariance of event authorization with the exception of revocations in the \emph{authorization safety} invariant.
Spec.~\ref{alg:acs-model} realizes authorization safety through collision-resistant event identifiers and timestamps, only revocations can lead to changes in discursive authorization due to the negated existential quantifier for revocations in $\mfn{authorizes}()$.

\begin{description}
    \item[Authorization Safety]
    For group state $G$ and new event $e_\mtyp{m}$,
    causal authorization of any event $e \in G$
    is invariant.
    Concurrent authorization is invariant with the exception of authorized revocations.
    \begin{align*}
    \forall G \in \mathbb{G}, e_m &\in \mathbb{E}, G' = G \cup \{ e_m \}, G.\mtd{valid}() \land G'\!.\mtd{valid}()\colon \\
    \forall e \in G&\colon \\
    & G.\mtd{pre}(e).\mtd{authorizes}(e) = G'\!.\mtd{pre}(e).\mtd{authorizes}(e) \\
    \land &e_m \notin \mathbb{E}_\mtyp{revoke} \lor \neg G'\!.\mtd{authorizes}(e_m) \\
    & \Rightarrow G.\mtd{authorizes}(e) = G'\!.\mtd{authorizes}(e) \\
    \end{align*}
    
\end{description}


In the end, access control on group state is about controlling the effect of logged events on queries, which we formalize as \emph{query safety}.
Discursive authorization is a necessary condition for an event to affect query results, and, per contrapositive, query results must be invariant under any discursively unauthorized event.
In Spec.~\ref{alg:acs-model}, query safety is realized by $\mfn{caps}()$, $\mfn{authorized}()$, and $\mfn{values}()$ queries acting on discursively authorized events only.

\begin{description}
    
    \item[Query Safety]
    For group state $G$ and new event $e_m$, if $e_m$ causes varying results of a query operation $G.\mtd{query}()$, it must be discursively authorized.
    \begin{align*}
    \forall G \in \mathbb{G}, & e_m \in \mathbb{E}, G' = G \cup \{ e_m \}, G.\mtd{valid}() \land G'\!.\mtd{valid}()\colon \\
    & G.\mtd{query}() \neq G'.\mtd{query}() \Rightarrow G.\mtd{authorizes}(e_m)
    \end{align*}

\end{description}

The expansion of the scope of effect of revocations to include concurrent events introduces non-monotonic logic~\cite{hellersteinKeepingCALMWhen2020}:
Depending on the reception order, correct entities may exhibit non-monotonic anomalies where they disagree on discursive authorization of backdated events at first, until eventually reaching consistency on $G$ and thereby consistency on discursive authorization~\cite{jacob_best_2025}.
To make backdating ineffective and revocations safe, we state the \emph{revocation safety} invariant to ensure that any backdated version of a discursively unauthorized event is also discursively unauthorized.
In Spec.~\ref{alg:acs-model}, the $\mfn{authorized}()$ query ensures revocation safety by looking for concurrent revocation events matching the grant of the event to authorize.

\begin{description}
    \item[Revocation Safety]
    For group state $G$ and invocation $v$, if $v$ is unauthorized at timestamp $G.\mtd{now}()$, then $v$ is also unauthorized at any earlier timestamp $T \subseteq G.\mtd{now}()$.
    
    \begin{align*}
    \forall G \in \mathbb{G}, v \in \mathbb{V}, T \subseteq G.\mtd{now}(), G.\mtd{valid}()\colon \\
    \neg G.\mtd{authorizes}(\mfn{event}(G.\mtd{now}(), v)) \\
    \Rightarrow \neg G.\mtd{authorizes}(\mfn{event}(T, v)) \\
    \end{align*}
\end{description}

As the semantics in Spec.~\ref{alg:acs-model} support our security invariants in any group state, i.e., including events of Byzantine entities, we call them Byzantine fault-tolerant.

\section{Verification}
\label{sec:formal-verification}

\begin{figure}
    \includegraphics[width=\linewidth]{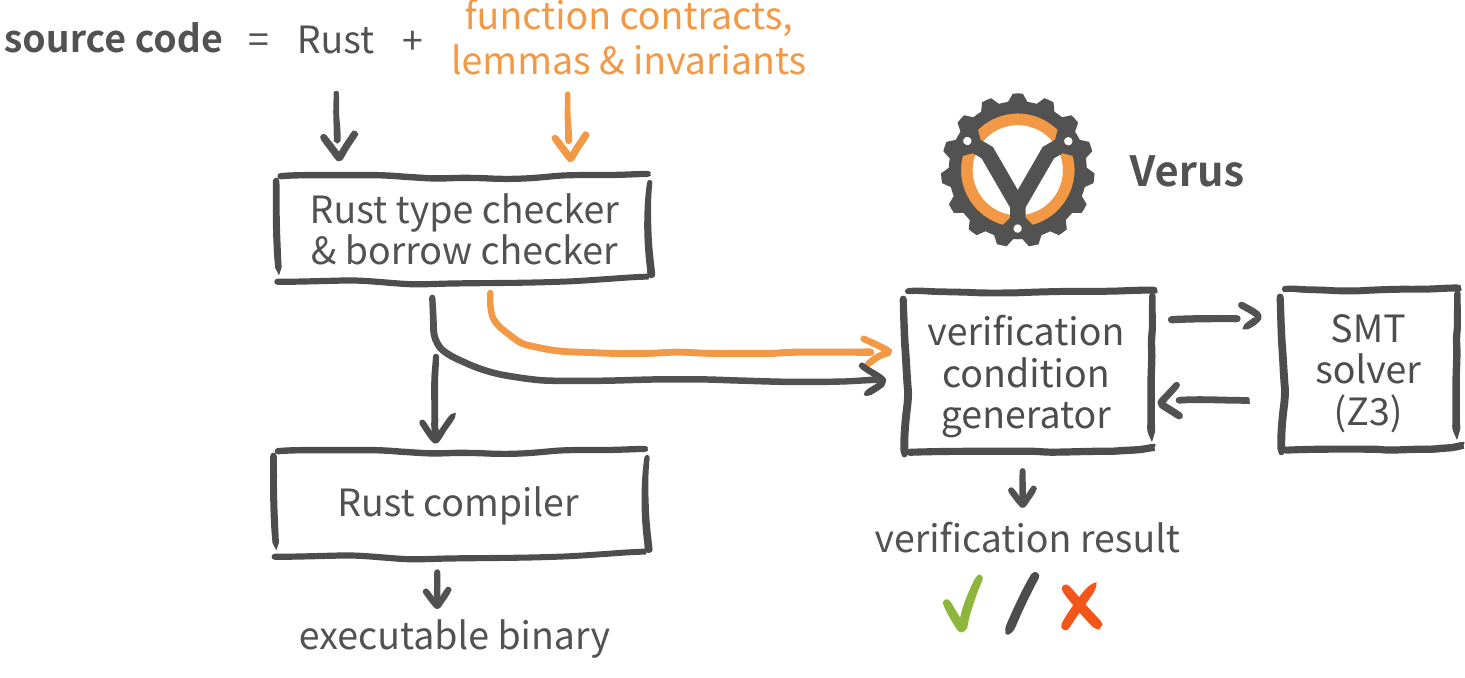}
    \caption{Verus extends Rust with formal verification at zero runtime cost by adding an SMT solver at compile time~\cite{lattuada_verus_2024}.}
    \label{fig:verus}
\end{figure}

As illustrated in \cref{fig:verus}, Verus~\cite{lattuada_verus_2024} is an SMT-based verification framework for Rust~\cite{matsakis_rust_2014} code.
Specification and verification add no runtime overhead, and Rust's ownership model rooted in linear type systems obviates the need for additional memory reasoning in separation logic.
An implementation is annotated with and verified against its formal specification in an extended Rust syntax (cf.~\cref{fig:matrix-vs-verus}).
Functions with the \texttt{exec} keyword are executed at runtime, i.e., make up the implementation.
Functions with the \texttt{spec} keyword make up the specification, connected with \texttt{exec} functions via \texttt{ensures} and \texttt{requires} conditions.
Specification code is called “ghost code”~\cite{lattuada_verus_2023} due to being present only during verification.

\subsection{Methodical Considerations}
\label{sec:method}

\emph{\textbf{Byzantine fault tolerance.}}
We abstract Byzantine fault tolerance of the underlying hash chronicle by assuming that authorization algorithms operate on a group object based on a valid chronicle.
However, valid chronicles still contain Byzantine behaviour, like events from equivocation or backdating.
Verifying the security invariants for authorization algorithms for any valid chronicle thereby also shows their Byzantine fault-tolerant implementation.


\emph{\textbf{Total functions}}
terminate and yield a result of the expected return type \cite{turnerTotalFunctionalProgramming2004}.
Proving that all employed functions are total even in Byzantine environment is necessary for latency and fault tolerance of local-first systems.
Verus verifies totality for \texttt{spec} functions and termination of any loops and recursive calls of \texttt{exec} functions.
Rust does not have an exception system, but functions can abort execution and unwind the call stack on severe error conditions via $\mfn{panic!}()$.
To guarantee that functions always yield a result, we must ensure that functions do not $\mfn{panic!}()$, which can be specified by the Verus keyword $\mvar{no\_unwind}$.
However, 
the \texttt{vstd} -- Verus' specification of Rust's \texttt{std} standard library -- does not yet provide $\mvar{no\_unwind}$ guarantees for any of the functions we used.
With a modified \texttt{vstd}, and under the assumption that \texttt{std} functions terminate and system calls for memory allocation and random number generation do not panic, we proved totality of all our functions.

\emph{\textbf{Pure functions}} deterministically derive the same output given the same input, and are free of side effects.
We verify strong convergence of query results by proving that all queries are total, pure functions of group state.
As \texttt{spec} functions in Verus are guaranteed to be pure, we can prove purity of query functions by specifying that the return value of the \texttt{exec} query is equal to its corresponding \texttt{spec} query.

\begin{figure}
    \begin{lstlisting}[columns=fullflexible,language=Rust]
spec fn authorizes(&self: GroupSpec, event: EventSpec) -> bool {
  match event.operation.operator() {
    Operator::Create => event == self.bot(),
    operator @ _ => {
      Self::has_cap(self.c.pre(event), event.subject, operator)
      ...
        \end{lstlisting}
        \vspace{-.75em}
        \begin{lstlisting}[columns=fullflexible,language=Rust]
exec fn log(&mut self, event: Event) -> (res: Option<EventID>)
  requires exists|id: EventID| !old(self)@.c.has_element(id), ...
  ensures res is Some <==> old(self)@.authorizes(event@), ... {
    if !self.authorizes(&event) { None } else {
      ...
        \end{lstlisting}
        \vspace{-.75em}
    \caption{
        Authorization rules formalized with Verus -- a function contract for the \texttt{exec}utable function \texttt{log} ensures that only authorized events are applied to the system state, as defined by the \texttt{spec}ification function \texttt{authorizes}.
    }
    \label{fig:matrix-vs-verus}
\end{figure}

\subsection{Authorization Algorithm Verification Case Study}
\label{sec:algorithms}

%

We successfully used Verus for formalizing and verifying a simplification of the group type semantics specification from \cref{sec:semantics} (source code cf.~\cite{mftf-source}).
We approached formalization in three steps with increasing complexity:
In the first step, capability grants are possible before, but not after group creation.
The second step enables capability grants after creation, but is still monotonic, i.e. does not support revocation.
In the third step, only revocations are supported after group creation, which introduces non-monotonic effects.



\emph{\textbf{1. Allow on creation.}}
This algorithm assumes that
grants are defined before creation and are immutable thereafter, i.e., the $\mtyp{grant}$ events logged before creation only grant the $\mtyp{assign}$ capability, but neither $\mtyp{grant}$ nor $\mtyp{revoke}$ capabilities.
This algorithm serves as baseline for other algorithms, focusing on authorization safety:
Because authorization is immutable, discursive authorization is equal to precursive authorization.
As only $\mtyp{assign}$ events are ever authorized, the conditions for verifying query safety also simplify substantially.

\emph{\textbf{2. Deny on creation, grant later.}}
This algorithm assumes that initial group setup only contains a single grant event, which grants the group creator the capability to grant entities the $\mtyp{assign}$ capability.
As no entity has the capability to revoke grants, capabilities evolve monotonically, and precursive and discursive authorization is still equivalent.
This step focuses on authorization and query safety, simplified by the equality between precursive and discursive authorization.


\emph{\textbf{3. Allow on creation, revoke later.}}
This algorithm assumes that all entities are granted the $\mtyp{assign}$ capability during group setup.
Additionally, the group creator is granted the $\mtyp{revoke}$ capability.
At first glance, a revoke-only access control system seems to be equally monotonic as a grant-only system: the capability set $\mfn{caps}()$ is now monotonically shrinking, instead of growing.
However, for any event $e$, the set of discursively authorized events that are concurrent to $e$, i.e., the set of events that may influence query results, can now both grow and shrink.
Consequently, the system behaviour is no longer monotonic:
In the light of new information, i.e., a revocation, discursive authorization of events can change, and with them previous beliefs on system state, requiring the revocation safety invariant.~\cite{hellersteinKeepingCALMWhen2020}.
By serving as the base case for non-monotonic behaviour, this algorithm allows to focus on the non-monotonic aspects of authorization and query safety invariants.

\section{Discussion}

\paragraph{Methodical Discussion}
For our verification, we needed to add one additional assumption missing in Verus' \texttt{vstd} specification of Rust's \texttt{std} standard library, namely that
cloning preserves the elements of a \texttt{HashSet}.
Like all other \texttt{vstd} specifications, this is true under the assumption that the \texttt{std} implementation is correct.

While many of our \texttt{exec} functions are straightforward translation of their \texttt{spec} counterpart to regular Rust code, Verus' limited support for the \texttt{Iterator} trait and combinator functions like \texttt{filter}, \texttt{map}, \texttt{any}, etc., required us to often implement such functionality with \texttt{for} loops instead.
Verifying \texttt{for} loops in place of what otherwise would have been a chain of iterator combinators was one of the most time-consuming tasks of the overall verification process.
We expect that future work will greatly benefit from iterator support in Verus currently in development.

\paragraph{Result Discussion}
Under the assumption of Byzantine fault-tolerant eventual consistency of the underlying hash chronicle~\cite{jacob_logical_2024}, our \emph{query results are eventually consistent} due to the proof of purity and totality of query functions.

In our verified algorithms, concurrently issued $\mtyp{assign}$ and $\mtyp{revoke}$ events are the only possible authorization-related concurrency conflict, avoiding mutual authorization revocations among equally-privileged entities.
This allowed us to directly specify and verify simplifications of mutate and query operations from Spec.~\ref{alg:acs-model} for our algorithms, which ensure \emph{query safety} via filtering for authorized events.
\emph{Precursive authorization safety} is directly translated into a type invariant of the group type.
\emph{Discursive authorization safety} and \emph{revocation safety} are ensured by the granted-present-revoke construction implemented in $\mfn{authorizes}()$.
The operation specifications, together with the definition of $\mfn{authorizes}()$, are strong enough to show that the invariant holds. 
\paragraph{Next Steps: Increasing Expressiveness.}
Although monotonic capabilities would drastically reduce the required complexity of local-first authorization, the ability to both grant and revoke capabilities allows entities to undo erroneous grants, and trust towards an entity might change over time.
The next step in expressiveness therefore is to \emph{support both $\mtyp{grant}$ and $\mtyp{revoke}$ events in a single algorithm}.
This necessitates to prove more complicated formulas with nested quantifiers to preserve the invariants. 
An extension with a $\mtyp{query}$ capability to be presented on invoking query operations as a way to represent group membership should then be sufficient for basic group management.

The next level is to support \emph{delegation}, i.e., granting of $\mtyp{grant}$ and $\mtyp{revoke}$ capabilities after group creation.
Delegations require handling or prevention of mutual, concurrent revocations, and will substantially increase verification complexity.
In addition, group members require the capability to self-revoke their $\mtyp{query}$ capability to be able to leave a group at their discretion.
Publicly-joinable groups even require granting the capability to self-grant the $\mtyp{query}$ capability to all system entities, so that they can join the group at their discretion by granting themselves the $\mtyp{query}$ capability.

In continuation, the semantics should allow for a form of \emph{sparse replication} of the group object without impeding its security guarantees, as well as support seamless \emph{changeover} from one group to the next for archival, garbage collection, or protocol upgrade purposes.
A verified implementation of these concepts would provide practical advantages that surpass the current state of the art.



\section{Conclusion}
\label{sec:conclusion}

We proposed a bottom-up approach to construct formally verified local-first authorization algorithms as a starting point towards the future goal of verifying algorithms with the expressiveness of real-world local-first systems such as Matrix and Keyhive.
We presented a first specification of capability-based semantics and invariants for local-first access control on a replicated data type for managing collaboration groups.
We also provided a proof of concept by formally verifying a simplified specification and implementation:
Assuming an eventually consistent chronicle, but allowing for backdating by Byzantine entities, we verified our security invariants.
Verus promises to be an accessible, future-proof tool for bridging the gap between current systems specifications and formal methods.
However, scaling up formalization and verification towards the level required for practical local-first access control remains a challenge.

\begin{acks}
This work was funded by the Helmholtz Pilot Program Core Informatics.
We like to thank Martin Kleppmann for his thoughts on local-first access control, as well as the anonymous PaPoC 2026 reviewers for their thoughtful and substantial reviews that will help us progress with our research.
\end{acks}

\newpage

\balance

\bibliographystyle{ACM-Reference-Format}
\bibliography{references}

@inproceedings{shapiroConflictFreeReplicatedData2011,
  title = {Conflict-{{Free Replicated Data Types}}},
  booktitle = {Stabilization, {{Safety}}, and {{Security}} of {{Distributed Systems}}},
  author = {Shapiro, Marc and Preguiça, Nuno and Baquero, Carlos and Zawirski, Marek},
  editor = {Défago, Xavier and Petit, Franck and Villain, Vincent},
  date = {2011},
  volume = {6976},
  pages = {386--400},
  publisher = {Springer Berlin Heidelberg},
  location = {Berlin, Heidelberg},
  doi = {10.1007/978-3-642-24550-3_29},
  url = {http://link.springer.com/10.1007/978-3-642-24550-3_29},
  urldate = {2024-01-02},
  isbn = {978-3-642-24550-3},
  langid = {english},
}

@online{zelenkaKeyhiveLocalfirstAccess,
  title = {Keyhive: {{Local-first}} Access Control},
  shorttitle = {Keyhive},
  author = {Zelenka, Brooklyn and Good, Alex},
  url = {https://www.inkandswitch.com/keyhive/notebook/},
  date = {2025},
  urldate = {2026-01-13},
  langid = {english},
}

@article{herlihyWaitfreeSynchronization1991,
  title = {Wait-Free Synchronization},
  author = {Herlihy, Maurice},
  date = {1991-01-01},
  journal = {ACM Transactions on Programming Languages and Systems},
  shortjournal = {ACM Trans. Program. Lang. Syst.},
  volume = {13},
  number = {1},
  pages = {124--149},
  issn = {0164-0925},
  doi = {10.1145/114005.102808},
  url = {https://dl.acm.org/doi/10.1145/114005.102808},
  urldate = {2024-06-04},
}

@online{kleppmannCritiqueCAPTheorem2015,
  title = {A Critique of the {{CAP}} Theorem},
  author = {Kleppmann, Martin},
  date = {2015-09-18},
  eprint = {1509.05393},
  eprinttype = {arXiv},
  eprintclass = {cs},
  doi = {10.48550/arXiv.1509.05393},
  url = {http://arxiv.org/abs/1509.05393},
  urldate = {2023-03-01},
  pubstate = {prepublished},
}

@online{sanjuanMerkleCRDTsMerkleDAGsMeet2020,
  title = {Merkle-{{CRDTs}}: {{Merkle-DAGs}} Meet {{CRDTs}}},
  shorttitle = {Merkle-{{CRDTs}}},
  author = {Sanjuan, Hector and Poyhtari, Samuli and Teixeira, Pedro and Psaras, Ioannis},
  date = {2020-04-27},
  eprint = {2004.00107},
  eprinttype = {arXiv},
  eprintclass = {cs},
  doi = {10.48550/arXiv.2004.00107},
  url = {http://arxiv.org/abs/2004.00107},
  urldate = {2023-12-11},
  pubstate = {prepublished},
}

@article{bailisCoordinationAvoidanceDatabase2014,
  title = {Coordination Avoidance in Database Systems},
  author = {Bailis, Peter and Fekete, Alan and Franklin, Michael J. and Ghodsi, Ali and Hellerstein, Joseph M. and Stoica, Ion},
  date = {2014-11-01},
  journaltitle = {Proceedings of the VLDB Endowment},
  shortjournal = {Proc. VLDB Endow.},
  volume = {8},
  number = {3},
  pages = {185--196},
  issn = {2150-8097},
  doi = {10.14778/2735508.2735509},
  url = {https://dl.acm.org/doi/10.14778/2735508.2735509},
  urldate = {2023-06-21},
}

@misc{mftf-source,
    title = {Towards System-Oriented Formal Verification of Local-First Access Control: Source Code},
    author = {Stuber, Johanna and Jacob, Florian},
    date = {2026},
    url = {https://codeberg.org/kit-dsn/towards-formal-verification-local-first-access-control},
}

@article{basin_it_2025,
    title = {It takes a village: bridging the gaps between current and formal specifications for protocols},
    volume = {68},
    issn = {0001-0782, 1557-7317},
    doi = {10.1145/3706572},
    language = {en},
    number = {8},
    journal = {Communications of the ACM},
    author = {Basin, David and Foster, Nate and McMillan, Kenneth L. and Namjoshi, Kedar S. and Nita-Rotaru, Cristina and Smith, Jonathan M. and Zave, Pamela and Zuck, Lenore D.},
    month = aug,
    year = {2025},
    pages = {50--61},
}

@online{element_software_gmbh_matrix_2025,
    title = {Matrix in {Germany}},
    url = {https://element.io/matrix-in-germany},
    language = {en-GB},
    year = {2025},
    urldate = {2025-08-28},
    author = {{Element Software GmbH}},
}

@misc{elementsoftwaregmbhNATONI2CEMessenger2024,
  title = {{NATO} {NI2CE} {Messenger} utilises the power of decentralised communication},
  author = {Element Software GmbH},
  date = {2024-10-03T08:50:58},
  url = {https://element.io/blog/nato-ni2ce-messenger-utilises-the-power-of-decentralised-communication/},
  organization = {Element Blog},
}

@misc{matrix_spec_v1.16,
    title = {Matrix Specification v1.16},
    url = {https://spec.matrix.org/v1.16/},
    language = {en},
    author = {{The Matrix.org Foundation CIC}},
    date = {2025-09-17},
    year = {2025}
}

@article{lattuada_verus_2023,
    title = {Verus: verifying {Rust} programs using linear ghost types},
    volume = {7},
    issn = {2475-1421},
    shorttitle = {Verus},
    doi = {10.1145/3586037},
    language = {en},
    number = {OOPSLA1},
    journal = {Proceedings of the ACM on Programming Languages},
    author = {Lattuada, Andrea and Hance, Travis and Cho, Chanhee and Brun, Matthias and Subasinghe, Isitha and Zhou, Yi and Howell, Jon and Parno, Bryan and Hawblitzel, Chris},
    month = apr,
    year = {2023},
    pages = {286--315},
}

@inproceedings{lattuada_verus_2024,
    title = {Verus: a practical foundation for systems verification},
    isbn = {979-8-4007-1251-7},
    shorttitle = {Verus},
    doi = {10.1145/3694715.3695952},
    language = {en},
    booktitle = {Proceedings of the {ACM} {SIGOPS} 30th {Symposium} on {Operating} {Systems} {Principles}},
    publisher = {ACM},
    author = {Lattuada, Andrea and Hance, Travis and Bosamiya, Jay and Brun, Matthias and Cho, Chanhee and LeBlanc, Hayley and Srinivasan, Pranav and Achermann, Reto and Chajed, Tej and Hawblitzel, Chris and Howell, Jon and Lorch, Jacob R. and Padon, Oded and Parno, Bryan},
    month = nov,
    year = {2024},
}

@inproceedings{kleppmann_local-first_2019,
    address = {Athens Greece},
    title = {Local-first software: you own your data, in spite of the cloud},
    isbn = {978-1-4503-6995-4},
    shorttitle = {Local-first software},
    doi = {10.1145/3359591.3359737},
    language = {en},
    booktitle = {Proceedings of the 2019 {ACM} {SIGPLAN} {International} {Symposium} on {New} {Ideas}, {New} {Paradigms}, and {Reflections} on {Programming} and {Software}},
    publisher = {ACM},
    author = {Kleppmann, Martin and Wiggins, Adam and Van Hardenberg, Peter and McGranaghan, Mark},
    month = oct,
    year = {2019},
    pages = {154--178},
}

@inproceedings{matsakis_rust_2014,
    title = {The {Rust} Language},
    isbn = {978-1-4503-3217-0},
    doi = {10.1145/2663171.2663188},
    language = {en},
    booktitle = {Proceedings of the 2014 {ACM} {SIGAda} annual conference on {High} integrity language technology},
    publisher = {ACM},
    author = {Matsakis, Nicholas D. and Klock, Felix S.},
    month = oct,
    year = {2014},
}

@inproceedings{weberAccessControlWeakly2016,
  title = {Access Control for Weakly Consistent Replicated Information Systems},
  booktitle = {Security and {{Trust Management}}},
  author = {Weber, Mathias and Bieniusa, Annette and Poetzsch-Heffter, Arnd},
  date = {2016},
  year = {2016},
  series = {Lecture {{Notes}} in {{Computer Science}}},
  doi = {10.1007/978-3-319-46598-2_6},
  isbn = {978-3-319-46598-2},
  langid = {english},
}

@phdthesis{rault_access_2024,
    title = {Access control mechanisms for collaborative systems without central authority},
    url = {https://hal.univ-lorraine.fr/tel-05079538},
    language = {en},
    school = {Université de Lorraine},
    author = {Rault, Pierre-Antoine},
    month = dec,
    year = {2024},
}

@inproceedings{jacob_best_2025,
    author       = {Jacob, Florian and Hartenstein, Hannes},
    year         = {2025},
    title        = {To the Best of Knowledge and Belief: On Eventually Consistent Access Control},
    pages        = {107--118},
    eventtitle   = {15th Association for Computing Machinery Conference on Data and Application Security and Privacy},
    eventtitleaddon = {ACM CODASPY 2025},
    eventdate    = {2025-06-04/2025-06-06},
    venue        = {Pittsburgh, PA, USA},
    booktitle    = {Proceedings of the 15th ACM Conference on Data and Application Security and Privacy},
    doi          = {10.1145/3714393.3726520},
    publisher    = {{Association for Computing Machinery (ACM)}},
    isbn         = {979-8-4007-1476-4},
    language     = {english}
}

@unpublished{kleppmannPresentFutureLocalfirst2024,
  title = {The Past, Present, and Future of Local-First},
  author = {Kleppmann, Martin},
  date = {2024-05-30},
  url = {https://martin.kleppmann.com/2024/05/30/local-first-conference.html},
  eventtitle = {Local-{{First Conference}} 2024},
  venue = {Berlin},
}

@inproceedings{raultAccessControlBased2023,
  title = {Access Control Based on {{CRDTs}} for Collaborative Distributed Applications},
  author = {Rault, Pierre-Antoine and Ignat, Claudia-Lavinia and Perrin, Olivier},
  date = {2023-11-01},
  url = {https://inria.hal.science/hal-04224855},
  urldate = {2024-01-27},
  eventtitle = {22nd {{IEEE International Conference}} on {{Trust}}, {{Security}} and {{Privacy}} in {{Computing}} and {{Communications}} ({{TrustCom-2023}})},
  langid = {english},
}

@online{p2pandacontributorsP2panda2026,
  author = {{p2panda Contributors}},
  date = {2026},
  url = {https://p2panda.org/},
  title = {p2panda: building blocks for peer-to-peer applications},
}

@online{automergecontributorsAutomergeVersionControl2026,
  title = {Automerge: Version Control for Your Data},
  author = {{Automerge Contributors}},
  date = {2026},
  url = {https://automerge.org/},
  langid = {english},
}

@article{jacobAnalysisMatrixEvent2021,
  title = {Analysis of the {{Matrix Event Graph}} Replicated Data Type},
  author = {Jacob, Florian and Beer, Carolin and Henze, Norbert and Hartenstein, Hannes},
  date = {2021},
  journal = {IEEE Access},
  volume = {9},
  pages = {28317--28333},
  issn = {2169-3536},
  doi = {10.1109/ACCESS.2021.3058576},
}

@article{gusmeroliCapabilitybasedSecurityApproach2013,
  title = {A Capability-Based Security Approach to Manage Access Control in the {{Internet}} of {{Things}}},
  author = {Gusmeroli, Sergio and Piccione, Salvatore and Rotondi, Domenico},
  date = {2013-09-01},
  journal = {Mathematical and Computer Modelling},
  shortjournal = {Mathematical and Computer Modelling},
  series = {The {{Measurement}} of {{Undesirable Outputs}}: {{Models Development}} and {{Empirical Analyses}} and {{Advances}} in Mobile, Ubiquitous and Cognitive Computing},
  volume = {58},
  number = {5},
  pages = {1189--1205},
  issn = {0895-7177},
  doi = {10.1016/j.mcm.2013.02.006},
  url = {https://www.sciencedirect.com/science/article/pii/S089571771300054X},
  urldate = {2026-01-23},
}

@article{sandhuAccessControlPrinciple1994,
  title = {Access Control: Principle and Practice},
  shorttitle = {Access Control},
  author = {Sandhu, R.S. and Samarati, P.},
  date = {1994-09},
  journal = {IEEE Communications Magazine},
  volume = {32},
  number = {9},
  pages = {40--48},
  issn = {1558-1896},
  doi = {10.1109/35.312842},
  url = {https://ieeexplore.ieee.org/document/312842},
  urldate = {2026-01-23},
}

@online{weidnerDesigningDataStructures2022,
  title = {Designing Data Structures for Collaborative Apps},
  author = {Weidner, Matthew},
  date = {2022-02-10},
  url = {https://mattweidner.com/2022/02/10/collaborative-data-design.html}
}

@article{gomes_verifying_2017,
    title = {Verifying strong eventual consistency in distributed systems},
    volume = {1},
    issn = {2475-1421},
    doi = {10.1145/3133933},
    number = {OOPSLA},
    journal = {Proceedings of the ACM on Programming Languages},
    author = {Gomes, Victor B. F. and Kleppmann, Martin and Mulligan, Dominic P. and Beresford, Alastair R.},
    month = oct,
    year = {2017},
    pages = {1--28},
}

@mastersthesis{da_design_2024,
    title = {Design and verification of Byzantine fault tolerant CRDTs},
    url = {https://github.com/TUM-DSE/research-work-archive/blob/main/archive/2024/winter/docs/msc_liangrun_da_design_and_verification_of_byzantine_fault_tolerant_crdts.pdf},
    school = {TUM},
    author = {Da, Liangrun},
    month = oct,
    year = {2024},
}

@inproceedings{kleppmann_papoc_keynote_2025,
    title = {Keynote: {Byzantine} eventual consistency and local-first access control},
    series = {12th Workshop on Principles and Practice of Consistency for Distributed Data (PaPoC '25)},
    url = {https://martin.kleppmann.com/2025/03/31/papoc-keynote-byzantine.html},
    date = {2025-03-31},
    author = {Kleppmann, Martin},
    publisher = {ACM},
}

@article{turnerTotalFunctionalProgramming2004,
  title = {Total Functional Programming},
  author = {Turner, D.},
  date = {2004-07-28},
  journal = {JUCS - Journal of Universal Computer Science},
  volume = {10},
  number = {7},
  pages = {751--768},
  publisher = {Journal of Universal Computer Science},
  issn = {0948-6968},
  doi = {10.3217/jucs-010-07-0751},
  url = {https://lib.jucs.org/article/28254/},
  urldate = {2024-09-19},
  issue = {7},
  langid = {english},
}

@article{hellersteinKeepingCALMWhen2020,
  title = {Keeping {{CALM}}: When Distributed Consistency Is Easy},
  shorttitle = {Keeping {{CALM}}},
  author = {Hellerstein, Joseph M. and Alvaro, Peter},
  date = {2020-08-21},
  journal = {Communications of the ACM},
  shortjournal = {Commun. ACM},
  volume = {63},
  number = {9},
  pages = {72--81},
  issn = {0001-0782},
  doi = {10.1145/3369736},
  url = {https://dl.acm.org/doi/10.1145/3369736},
  urldate = {2023-05-10},
}

@inproceedings{jacob_logical_2024,
    address = {Athens Greece},
    title = {Logical {Clocks} and {Monotonicity} for {Byzantine}-{Tolerant} {Replicated} {Data} {Types}},
    isbn = {979-8-4007-0544-1},
    doi = {10.1145/3642976.3653034},
    language = {en},
    booktitle = {Proceedings of the 11th {Workshop} on {Principles} and {Practice} of {Consistency} for {Distributed} {Data}},
    publisher = {ACM},
    author = {Jacob, Florian and Hartenstein, Hannes},
    month = apr,
    year = {2024},
}

@misc{almeida_blocklace_2025,
    title = {The {Blocklace}: {A} {Byzantine}-repelling and {Universal} {Conflict}-free {Replicated} {Data} {Type}},
    shorttitle = {The {Blocklace}},
    doi = {10.48550/arXiv.2402.08068},
    publisher = {arXiv},
    author = {Almeida, Paulo Sérgio and Shapiro, Ehud},
    month = jan,
    year = {2025},
    note = {arXiv:2402.08068 [cs]},
}

@inproceedings{kleppmann_making_2022,
    address = {New York, NY, USA},
    series = {{PaPoC} '22},
    title = {Making {CRDTs} {Byzantine} fault tolerant},
    isbn = {978-1-4503-9256-3},
    doi = {10.1145/3517209.3524042},
    booktitle = {Proceedings of the 9th {Workshop} on {Principles} and {Practice} of {Consistency} for {Distributed} {Data}},
    publisher = {Association for Computing Machinery},
    author = {Kleppmann, Martin},
    month = apr,
    year = {2022},
    pages = {8--15},
}

@misc{dougal_hydra_2025,
    title = {Project {Hydra}: improving state resolution in {Matrix}},
    shorttitle = {Project {Hydra}},
    url = {https://matrix.org/blog/2025/08/project-hydra-improving-state-res/},
    language = {en},
    date = {2025-08-14},
    year = {2025},
    month = {Aug},
    author = {Dougal, Kegan},
}

@misc{matrix_rooms_v12,
    title = {Room {Version} 12},
    url = {https://spec.matrix.org/v1.16/rooms/v12/},
    language = {en},
    date = {2025-09-17},
    journal = {Matrix Specification},
    author = {{The Matrix.org Foundation CIC}},
    year = {2025}
}

@article{almeida_approaches_2024,
    title = {Approaches to conflict-free replicated data types},
    volume = {57},
    issn = {0360-0300},
    doi = {10.1145/3695249},
    number = {2},
    journal = {ACM Comput. Surv.},
    author = {Almeida, Paulo Sérgio},
    month = nov,
    year = {2024},
    pages = {51:1--51:36},
}

\end{document}